 \documentclass[final,3p,times,twocolumn]{elsarticle}
 \usepackage{graphicx}
\usepackage{amssymb}
\def\Vec#1{\mbox{\boldmath $#1$}}

\journal{Physica C}

\begin{document}

\begin{frontmatter}




\title{
Analysis of 
field-angle dependent specific heat in unconventional superconductors:
a comparison between Doppler-shift method and Kramer-Pesch approximation
}

\author[Nano,CREST]{Nobuhiko Hayashi}
%\ead{n-hayashi@21c.osakafu-u.ac.jp} 
\author[Tokyo,TRIP]{Yuki Nagai}
\author[Osakafu]{Yoichi Higashi}

\address[Nano]{
Nanoscience and Nanotechnology Research Center (N2RC),
Osaka Prefecture University, 1-2 Gakuen-cho, Sakai 599-8570, Japan
}
\address[CREST]{
CREST(JST), 4-1-8 Honcho, Kawaguchi, Saitama 332-0012, Japan
}

\address[Tokyo]{
Department of Physics, University of Tokyo, Tokyo 113-0033, Japan
}
\address[TRIP]{JST, TRIP, Chiyoda, Tokyo, 102-0075, Japan}

\address[Osakafu]{
Department of Mathematical Sciences,
Osaka Prefecture University, 1-1 Gakuen-cho, Sakai 599-8531, Japan
}

\begin{abstract}
We theoretically discuss the magnetic-field-angle dependence of the zero-energy density of states (ZEDOS) in superconductors.
Point-node and line-node superconducting gaps on spherical and cylindrical Fermi surfaces are considered.
The Doppler-shift (DS) method and the Kramer-Pesch approximation (KPA)
are used to calculate the ZEDOS.
Numerical results show that consequences of  the DS method are corrected by the KPA.
\end{abstract}


\end{frontmatter}


%\section{Introduction}
In the last decade,
experimental techniques for
the magnetic-field-angle dependent measurements of
the thermal conductivity and specific heat have been developed
to investigate the superconducting gap anisotropy in various materials \cite{Matsuda,Sakakibara}.
The Doppler-shift (DS) method \cite{Volovik} has been frequently utilized to
analyze such measurements \cite{Vekhter,Won}.
Theoretical development has been in progress
so far \cite{Miranovic,Kusunose,Udagawa,Vorontsov,Alrub,Boyd,Nagai2007,Nagai2008,Nagai2009}.
Recently, we have developed a new method called the Kramer-Pesch approximation (KPA) \cite{Nagai2008,Nagai2006},
where the contribution of a vortex core neglected in the DS method is taken into account \cite{Nagai2008}.
Then, the KPA may be a useful tool that enables a more quantitative analysis.

In this paper,
we show several numerical results for the zero-energy density of states (ZEDOS)
obtained by the DS method and the KPA
to demonstrate how much the KPA improves quantitative results of the DS method.
The specific heat over the temperature, $C(T)/T$, is proportional to the ZEDOS
in the low temperature limit $T \to 0$.

%%%%%%%%

The expression for the density of states (DOS) $N(E)$ in the DS method is given as \cite{Volovik,Vekhter,Nagai2007}
\begin{eqnarray}
N(E)
\propto
{\rm Re}
\Biggl\langle
\int dS_{\rm F}
\frac{|E-\delta E|}{\sqrt{(E-\delta E)^2 - |\Delta|^2} }
\Biggr\rangle_{\rm DS},
\label{eq:1}
\end{eqnarray}
where
$\delta E=m {\Vec v}_{\rm F} \cdot {\Vec v}_{\rm s}$
is the DS energy.
$m$, ${\Vec v}_{\rm F}$, and ${\Vec v}_{\rm s}$
are the electron mass, the Fermi velocity, and the circulating superfluid velocity around a vortex, respectively.
The field-angle dependence is taken into account via ${\Vec v}_{\rm s}$,
which is perpendicular to the magnetic field ${\Vec H}$ (i.e., ${\Vec v}_{\rm s} \perp {\Vec H}$).
Around a single vortex, $|{\Vec v}_{\rm s}|=\hbar/2mr$ ($r$ is the radial distance from the vortex center).
%The brackets
%$\langle \cdots \rangle_{\rm DS}$
%mean the spatial integration around the vortex,
%\begin{eqnarray}
$
%\bigl
\langle \cdots
%\bigr
\rangle_{\rm DS}
=
 \int_{\xi_0}^{r_{\rm a}} r dr \int_0^{2 \pi} d \alpha  \cdots$,
%\end{eqnarray}
is the spatial integration around the vortex
in the cylindrical coordinates $(r,\alpha,z)$ with
${\hat z} \parallel {\Vec H}$.
Here,
$\xi_0$ is the coherence length and
$r_a$ is the cutoff length
with $r_a/\xi_0 = \sqrt{H_{c2}/H}$,
($H_{c2} \equiv \Phi_0 / \pi \xi_0^{2}$, $\Phi_0 = \pi r_a^2 H$,
and $\Phi_0$ is the flux quantum).
%%%
%In Eq.~(\ref{eq:1}), 
$d S_{\rm F}$ is an area element on the Fermi surface (FS)
[e.g.,
$d S_{\rm F}=k_{\rm F}^2 \sin\theta d\phi d\theta$
for a spherical FS
in the spherical coordinates $(k,\phi,\theta)$,
and
$d S_{\rm F}=k_{{\rm F}ab} d\phi dk_c$
for a cylindrical FS
in the cylindrical coordinates $(k_{ab},\phi,k_c)$].
The pair potential is $\Delta \equiv \Delta_0 \Lambda({\Vec k}_{\rm F})$,
where $\Delta_0$
is the maximum gap amplitude
and
$\Lambda({\Vec k}_{\rm F})$
represents the gap anisotropy on the FS.
$\Lambda$ and ${\Vec v}_{\rm F}$ are functions of the position ${\Vec k}_{\rm F}$ on the FS.

%%%%%%%%

On the other hand,
the DOS in the KPA is given as~\cite{Nagai2008}
\begin{eqnarray}
N(E)
=
\frac{v_{\rm F0} \eta}
     {2 \pi^2 \xi_0}
\Biggl{\langle} 
\int 
\frac{ d S_{\rm F}  }
     { |\Vec{v}_{\rm F}|  }
\frac{
  \lambda
  \bigl[ \cosh
        (x/\xi_0)
  \bigr]^{-2 \lambda / \pi h} 
 }{(E-E_y)^2 + \eta^2} 
\Biggl{\rangle}_{\rm KPA}.
\label{eq:dos}
\end{eqnarray}
%%%%%%%%%%%%%%%%%%%%%%%%%%%%%%
%the pair potential is $\Delta \equiv \Delta_0 \Lambda({\Vec k}_{\rm F}) \tanh(r/\xi_0) \exp(i\alpha)$
%around a vortex,
Here,
$\lambda = |\Lambda|$,
and
$\langle \cdots \rangle_{\rm KPA}
\equiv \int_0^{r_a} r dr \int_0^{2 \pi} \cdots d \alpha / \pi r_a^2$
is the real-space average around a vortex
in the cylindrical coordinates $(r,\alpha,z)$ with
${\hat z} \parallel {\Vec H}$.
$x=r\cos(\alpha -\theta_v)$,
$y=r\sin(\alpha -\theta_v)$,
and
$E_y = \Delta_0 \lambda^2 y /  \xi_0 h$.
$\theta_v({\Vec k}_{\rm F},\alpha_{\rm M}, \theta_{\rm M})$ 
is the angle of ${\Vec v}_{\rm F \perp}$ in the plane of $z=0$,
where $\alpha$ and $\theta_v$ are measured from a common axis \cite{Nagai2007,Nagai2006}.
$\Vec{v}_{\rm F \perp}$
is the vector component of ${\Vec v}_{\rm F} ({\Vec k}_{\rm F})$
projected onto the plane normal to
${\hat {\Vec H}}=(\alpha_{\rm M}, \theta_{\rm M})$.
Here,
the azimuthal (polar) angle of ${\Vec H}$ is $\alpha_{\rm M}$ ($\theta_{\rm M}$)
in a spherical coordinate frame fixed to crystal axes.
$|{\Vec v}_{\rm F \perp}({\Vec k}_{\rm F},\alpha_{\rm M}, \theta_{\rm M})|
\equiv v_{\rm F0}(\alpha_{\rm M}, \theta_{\rm M}) h({\Vec k}_{\rm F},\alpha_{\rm M}, \theta_{\rm M})$
and 
$v_{\rm F0}$ is the FS average of $|{\Vec v}_{\rm F \perp}|$ 
represented as
\cite{Nagai2007},
%\begin{eqnarray}
$v_{\rm F0}
=
\int dS_{\rm F} |{\Vec v}_{\rm F \perp}| / \int dS_{\rm F}$.
%\frac{\int dS_{\rm F} |{\Vec v}_{\rm F \perp}|}{\int dS_{\rm F}}.
%\end{eqnarray}
The coherence length is expressed as
$\xi_0 = \hbar v_{{\rm F}0}/ \pi \Delta_0$.
The smearing factor is set as $\eta=0.05\Delta_0$.

%%%%%%%%

We numerically calculate the ZEDOS $N(0)$
for the following four combinations: the line-node
or the point-node gap on the cylindrical or the spherical
FS.
Here, 
the line-node gap is
the $d$-wave pairing $\Lambda=\cos2\phi\sin^2\theta$
for the spherical FS
and is
$\Lambda=\cos2\phi$
for the cylindrical FS.
The point-node gap is the $s+g$ wave
$\Lambda= (1+\sin^4 \theta \cos 4 \phi)/2$
for the spherical FS
and is
$\Lambda= (1+\cos^4 (k_c/2) \cos 4 \phi)/2$
for the cylindrical FS.

%%%%%%%%

We show the results in Figs.~1 and 2.
As shown in Fig.~1(a) [the line-node gap on the spherical FS],
a cusp-like minimum does not appear
and the oscillation amplitude is of the order 3.5 \% for the KPA, coinciding well with
a full numerical result in Ref.~\cite{Miranovic}.
For the DS result, however, the amplitude is of the order 4.9 \% with less coincidence.
As seen in all figures,
the DS method is liable to give rise to an inflated oscillation amplitude.
While a cusp-like minimum appears both for the DS method and the KPA in Fig.~2(a)  [the line-node gap on the cylindrical FS],
only the DS method gives rise to a relatively cusp-like structure in the case of the point-node gap on the cylindrical FS as noticed in Fig.~2(b),
which we mentioned in Ref.~\cite{Nagai2008}.
The KPA is a perturbation theory, and the DS method corresponds to its zeroth order one with a spatially constant gap~\cite{Nagai2008}.
Therefore, that discrepancy in Fig.~2(b) signifies an improvement due to the KPA.

Both the DS method and the KPA are theories valid in low-field and low-$T$ region.
Taking into account a contribution of a vortex core \cite{Nagai2008},
the KPA yields more quantitative results with the same computational load compared to the DS method.

%%%%%%%%%
\begin{figure}
%\includegraphics[width = 8.5cm]{Fig2}% Here is how to import EPS art
%  \begin{center}
%\includegraphics[width = 50mm]{pointnodes_sphere.eps}
%\includegraphics[width = 50mm]{linenodes_sphere.eps}
\hspace{-1cm}
    \begin{tabular}{p{35mm}p{35mm}}
      \resizebox{56mm}{!}{\includegraphics{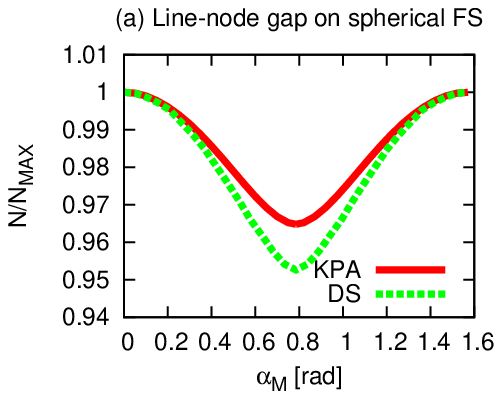}} &
      \resizebox{56mm}{!}{\includegraphics{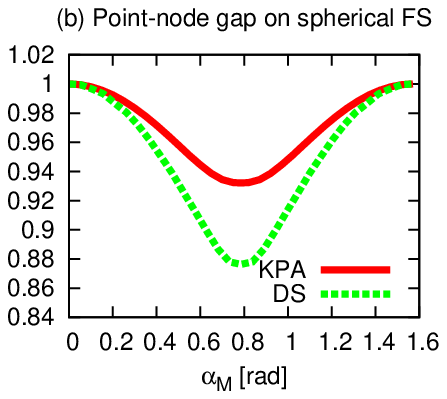}} 
    \end{tabular}
\caption{\label{fig:fig1}
Azimuthal field-angle $\alpha_{\rm M}$ dependence of the ZEDOS for the spherical FS.
The polar angle $\theta_{\rm M}=\pi/2$.
The cutoff length $r_a = 7\xi_0$. 
}
%  \end{center}
\end{figure}
%%%%%%%%%

%%%%%%%%%
\begin{figure}
  \begin{center}
\hspace{-1.8cm}
    \begin{tabular}{p{35mm}p{35mm}}
      \resizebox{56mm}{!}{\includegraphics{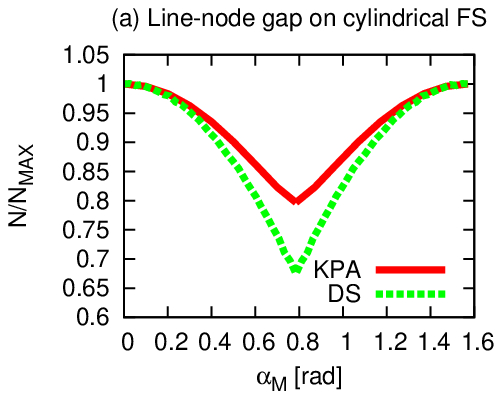}} &
      \resizebox{56mm}{!}{\includegraphics{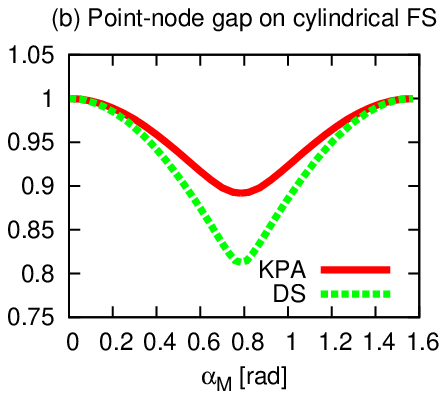}} 
    \end{tabular}
\caption{\label{fig:fig2}
Azimuthal field-angle $\alpha_{\rm M}$ dependence of the ZEDOS for the cylindrical FS.
The polar angle $\theta_{\rm M}=\pi/2$.
The cutoff length $r_a = 7\xi_0$.
}
  \end{center}
\end{figure}
%%%%%%%%%

%\section*{Acknowledgments}
One of us (Y.N.) acknowledges support 
   by Grant-in-Aid for JSPS Fellows (204840). 
%%%%%%%%

%\appendix*
%\section{}   % <== necessary!
%%%
%\section{}



\end{document}